# Sorption of Eu(III) on Attapulgite Studied by Batch, XPS and EXAFS Techniques

Q.H. FAN[†,‡,#], X.L. TAN[†,#], J.X. LI[†], X.K. WANG[†*], W.S. WU[‡*], G. Montavon[&*]

Key Laboratory of Novel Thin Film Solar Cells, Institute of Plasma Physics, Chinese Academy of Sciences, P.O. Box 1126, Hefei, 230031, P.R. China; Radiochemistry Laboratory, Lanzhou University, Lanzhou, 730000, P.R. China; and Laboratory SUBATECH, UMR 6457 Ecole des Mines de Nantes/IN2P3-CNRS/Université de Nantes, 4 rue A. Kastler, BP 20722, 44307 Nantes cedex 03, France

*: Corresponding author. Tel: +86-551-5592788; Fax: +86-551-5591310. Email: xkwang@ipp.ac.cn (X.K. Wang); wuws@lzu.edu.cn (W.S. Wu). montavon@subatech.in2p3.fr (G. Montavon). #: same contribution to the paper. †: Institute of Plasma Physics; ‡: Lanzhou University; &: Laboratory SUBATECH.

The effects of pH, ionic strength and temperature on sorption of Eu(III) on attapulgite were investigated in the presence and absence of fulvic acid (FA) and humic acid (HA). The results indicated that the sorption of Eu(III) on attapulgite was strongly dependent on pH and ionic strength, and independent of temperature. In the presence of FA/HA, Eu(III) sorption was enhanced at pH < 4, decreased at pH range of 4 - 6, and then increased again at pH > 7. The X-ray photoelectron spectroscopy (XPS) analysis suggested that the sorption of Eu(III) might be expressed as $\equiv X_3Eu^0$ $\equiv S^wOHEu^{3+}$ and $\equiv SOEu-OOC-/HA$ in the ternary Eu/HA/attapulgite system. The extended X-ray absorption fine structure (EXAFS) analysis of Eu-HA complexes indicated that the distances of d(Eu-O) decreased from 2.451 to 2.360 Å with increasing pH from 1.76 to 9.50, whereas the coordination number (N) decreased from ~9.94 to ~8.56. Different complexation species were also found for the different addition sequences of HA and Eu(III) to attapulgite suspension. The results are important to understand the influence of humic substances on Eu(III) behavior in the natural environment.

## Introduction

In the context of safety of nuclear waste repositories as well as for the assessment of radionuclide mobility in the environment, the interaction between actinides and humic substances (HS) has been the subject of various studies. Europium is usually taken as a homologue for trivalent actinides because the ionic radius of Eu(III) is almost the same for the trivalent lanthanides and actinides. Therein, the sorption of Eu(III) at the solid-water interface is important for the performance assessment of nuclear waste repository *(1)*. The results of Eu(III) sorption on kaolinite and montmorillonite indicated that a unique inner-sphere complex ($\equiv$AlOEu$^{2+}$) linked to the aluminol sites was assumed at the edge of minerals, however, a second exchangeable outer-sphere complex for montmorillonite probably presented in an interlayer *(2)*. The sorption edge of Eu(III) on Ca-montmorillonite and Na-illite could be quantitatively modeled in pH range ~3 to ~10 using cation exchange reaction for Eu$^{3+}$/Na$^{+}$ and Eu$^{3+}$/Ca$^{2+}$ and three surface complexation reactions on the strong sorption sites forming $\equiv$S$^s$OEu$^{2+}$, $\equiv$S$^s$OEuOH$^{+}$ and $\equiv$S$^s$OEu(OH)$_2^0$ inner-sphere complexes *(3)*. The quantitative description of the ternary Eu/HS/mineral systems show that Eu(III) sorption is mainly governed by the behavior of HS, especially when strong complexes are formed between HS and Eu(III) ions. The presence of HS increases the sorption of Eu(III) at acidic pH values but reduces Eu(III) sorption at high pH values, which is attributed to the sorption of HS on the mineral surface followed by the interaction of Eu(III) with surface adsorbed HS at low pH values, whereas the formation of soluble Eu-HS complexes stabilizes Eu(III) ion in aqueous solution at high pH values *(4, 5)*. Spectroscopy analysis (such as extended X-ray absorption fine structure (EXAFS) spectroscopy, time resolved laser fluorescence spectroscopy (TRLFS) and X-ray photoelectron spectroscopy (XPS)) can offer structural information at molecular level. The EXAFS study of the reactions between iron and fulvic acid (FA) in acid aqueous solution indicated that iron was octahedrally configured with inner-sphere Fe-O interaction at 1.98 - 2.10 Å depending on the oxidation state iron *(6)*. Stumpf et al. *(7)* studied the incorporation of Eu(III) on hydrotalcite using TRLFS and EXAFS, and found that minor part of Eu(III) was inner-sphere adsorbed to the mineral surface, while the dominating Eu/hydrotalcite species consisted of Eu(III) that was incorporated into the hydrotalcite lattice. The interaction of Am(III) with 6-line-ferrihydrite was investigated by EXAFS, and found that Am(III) was adsorbed as a bidentate corner-sharing species at low pH values (pH 5.5) as well at higher pH values (pH 8.0) *(8)*.

Attapulgite, which is a hydrated magnesium aluminium silicate present in nature as fibrillar mineral, generally has three kinds of water at room temperature: (i) free water; (ii) zeolite water; and (iii) crystalline water. In addition, some isomorphic substitutions in the tetrahedral layer, such as $Al^{3+}$ for $Si^{4+}$, develop negatively charged sorption sites to electro-statically adsorb cation ions (9). Its ideal structure is shown in Figure SI-1. The special structure and surface properties of attapulgite make it a very suitable sorbent in the removal of heavy metal ions. However, to the best of our knowledge, the studies of Eu(III) sorption on attapulgite in the absence and presence of HS is not available, especially the spectroscopic study of Eu(III) structure at molecular level.

The present work aims to study the effects of pH, ionic strength, temperature and fulvic acid (FA) / humic acid (HA) on Eu(III) sorption to attapulgite. The species and mechanism of Eu(III) sorption to attapulgite are modeled and analyzed using FITEQL 3.2 code and XPS. Effect of HA/Eu(III) addition sequences on Eu(III) sorption to HA-attapulgite hybrids are also investigated. The interaction of Eu(III) with HA bound attapulgite is investigated by using Eu $L_{III}$-edge EXAFS to determine the local structure around the select element atom, which is very important to clarify the controversy of the influence of the HA/Eu(III) addition sequences on the molecular structures of Eu(III) adsorbed on HA-attapulgite hybrids. The combination of two complementary spectroscopic methods (i.e., XPS and EXAFS) and FITEQL 3.2 code allow us to understand and to quantify the adsorbed Eu(III) species and sorption mechanism.

**Experimental Section**

**Materials.** Eu(III) stock solution was prepared from $Eu_2O_3$ after dissolution, evaporation and redissolution in $10^{-3}$ mol/L perchloric acid. The sample of raw attapulgite was achieved from Kaidi Co. (Gansu, China). It was changed into Na-type (i.e., Na-attapulgite) and used in the experiments. The $N_2$-BET surface area of Na-attapulgite was 100.7 $m^2/g$. The chemical component is listed in Table SI-1, and the XRD patterns and FTIR spectra are shown Figures SI-2 and SI-3. The point of zero charge (PZC) is found to be 6.04 by potentiometric titration (Figure SI-4) (*10*).

FA and HA samples were extracted from the soil of Hua-Jia county (Gansu province, China), and were characterized in detail (*11*).

**Sorption Experiments.** The sorption experiments were carried out with 0.1 g/L attapulgite, $1.0 \times 10^{-6}$ mol/L Eu(III) and ionic strength (0.01 and 0.1 mol/L $NaClO_4$) at $T = 20 \pm 1$ °C under ambient conditions in the presence and absence of 10 mg/L FA or HA using batch technique. After sorption equilibrium, the pH values at higher temperatures were determined using pH meters by temperature compensation to eliminate the error of the measurements. Detailed experimental process is shown in Supporting Information (SI) (S2.2).

**Sample Preparation for XPS and EXAFS Analysis.** Detailed processes for the preparation are shown in SI (S2.3).

**XPS Analysis.** The XPS spectra were recorded on powders with a thermo ESCALAB 250 spectrometer using an Al K*a* monochromatized source and a multidetection analyzer under a $10^{-8}$ Pa residual pressure. Surface charging effects were corrected with C 1s peak at 284.6 eV as a reference. Shirley background correction and Gaussian-Lorentzian fitting were used to transform peak areas to total intensities.

**EXAFS Analysis.** Eu $L_{III}$-edge X-ray absorption spectra at 6976.9 eV were recorded at the National Synchrotron Radiation Laboratory (NSRL, China). Detailed descriptions of EXAFS analysis are listed in SI (S3.6).

## Results and Discussion

**Effect of ionic strength.** Figure 1A shows the removal percentage of Eu(III) as a function of pH in 0.01 and 0.1 mol/L NaClO$_4$ solutions, respectively. The removal percentage of Eu(III) on attapulgite is strongly dependent on pH and ionic strength (effect of NaClO$_4$ concentration on Eu(III) sorption is shown in Figure SI-6).

In 0.01 mol/L NaClO$_4$ solution, the sorption percentage of Eu(III) increases from ~3% to ~90% in the pH range from 1 to 5, and then remains constant with increasing pH values. However, in 0.1 mol/L NaClO$_4$ solution, the sorption percentage of Eu(III) increases from ~0.4% to ~90% in the pH range from 1.7 to 8. The surface charge of attapulgite is positive at pH < pH$_{pzc}$ (pH$_{pzc}$ = 6.04, Figure SI-4) due to an excess of protons on the surface and is negative at pH > pH$_{pzc}$. Eu(III) sorption is mainly dominated by the exchange of protons and Na$^+$ ions with Eu(III) at pH < pH$_{pzc}$; whereas the increasing negative charge enhances Eu(III) sorption until a plateau is reached, which indicates the main sorption mechanism is surface complexation at pH > pH$_{pzc}$. In the absence of attapulgite, the Eu(OH)$_3$ precipitation curve at C[Eu(III)] = 1.0 × 10$^{-6}$ mol/L is also shown in Figure 1A. The precipitation constant of Eu(OH)$_3$(s) ($K_{sp}$) is 3.4 × 10$^{-22}$. Europium ions begin to form precipitation at pH ~ 8.5 in the absence of attapulgite. It is clear that 90% Eu(III) is adsorbed to attapulgite at pH < 7. Thereby the sorption of Eu(III) on attapulgite at pH below 7 was not attributed to the surface precipitation of Eu(OH)$_3$, whereas surface precipitation of

Eu(OH)$_3$ may play role at pH values above 7 (*12*).

At pH < 8, the sorption curve of Eu(III) is shifted to the left in 0.01 mol/L NaClO$_4$ solution as compared to that of Eu(III) in 0.1 mol/L NaClO$_4$ solution. It is consistent with the results of the sorption of Eu(III) and Nd(III) on Marblehead illite (*13*). The results of Ni(II) sorption to Na-attapulgite indicated outer-sphere complexation or ion exchange might be the main sorption mechanism of Ni(II) sorption to Na-attapulgite at pH < 8, whereas the uptake of Ni(II) at pH > 8 was mainly dominated by inner-sphere complexation (*14*). The strong influence of pH and ionic strength on Eu(III) sorption can be explained by the contribution of sorption sites (≡SOH) exhibiting amphoteric properties at the surface of attapulgite (the amphoteric properties are listed in SI (Figure SI-5 and Table SI-2)). Two different sorption mechanisms are proposed: (1) exchange reaction with little or no dependency on pH, which dominates sorption at low pH; and (2) strong pH-dependent surface complexation reaction, which dominates sorption at neutral to alkaline conditions (*2, 13, 15*).

**Effect of Eu(III) concentration.** Figure 1B shows Eu(III) sorption on attapulgite at different initial Eu(III) concentrations as a function of pH values. The uptake curves show a typical "sorption edge", namely, the percentage of uptake increases from practically zero to about 90% over a range of more than three pH unites. However, 100% sorption is not reached, which can be interpreted in terms of the formation of water soluble carbonate complexes of Eu(III) ions (*16*). As expected, the pH-edge shifts to higher pH values at higher Eu(III) concentration, which is consistent with the results of Eu(III) sorption on natural hematite (17).

**Effect of temperature.** Figure 1C indicates that there is slight increase with increasing temperature at high pH, whereas Eu(III) sorption is independent of temperature at low pH. In order to adsorb cations to solid surface, cations are some extent denuded their hydration sheath in aqueous; this process requires energy ($Q_1$). If this energy of dehydration ($Q_1$) exceeds the exothermicity of cations attaching to the surface ($Q_2$) (i.e., $Q_1 > Q_2$), the sorption is an endothermicity process. If $Q_1 = Q_2$, then the sorption is independent of temperature. The results are consistent with the sorption of Eu(III) on smectite and on kaolinite (18, 19). Bauer et al. (18) reported the temperature (25 – 80 $^oC$) had no effect on the sorption of Eu(III) to smectite, and found ulteriorly the sorption of Eu(III) on smectite as outer-sphere complexation at low pH values and as inner-sphere complexation at pH > 5.5. Stumpf et al. (19) also reported similar sorption mechanism of Cm(III) on kaolinite and smectite. However, Tertre et al. (2) studied the sorption of Eu(III) to kaolinite and Na-montmorillonite as a function of temperatures (20 – 150 $^oC$), and found the increasing of temperature leaded to a shift of the sorption edge toward lower pH, and the shift was more pronounced for Na-montmorillonite than for kaolinite. The results suggest that the sorption of Eu(III) on minerals is dependent on the nature of the minerals.

**Effect of FA and HA.** The sorption of Eu(III) to attapulgite in the presence / absence of FA/HA as a function of pH is shown in Figure 1D. The sorption is enhanced slightly at pH < 4, decreased with increasing pH at pH 4-7, and then increased again with increasing pH at pH > 7 in the presence of FA. Similar results of Eu(III) sorption in the presence of HA is also observed. This may be attributed to the species of Eu(III) at different pH values. The relative species of Eu(III) as a function of pH in absence and presence of HA are shown in Figure SI-7. In the presence of HA, $Eu^{3+}$(HA) is the main species at pH < 6; $Eu^{3+}$(HA), $Eu(OH)^{2+}$(HA) and $Eu(CO_3)^+$(HA) are the dominated species at pH 6-10; and $Eu(CO_3)_3^{3-}$ becomes the main species at pH > 10 (Figure SI-7). Fairhurst and Warwick (20) investigated the influence of HA on Eu(III) sorption to mineral and found that $Eu^{3+}$(HA) complexes were formed at low pH which was adsorbed by mineral to the same extent as HA, leading to the enhanced Eu(III) sorption. The sorption of HS on attapulgite decreases with increasing pH values at pH > $pH_{pzc}$ because of the electrostatic repulsion between the negative charged surface of attapulgite and HS. Less Eu(III) is adsorbed to HS bound attapulgite due to the formation of soluble $Eu^{3+}$(HS) complexes in solution at pH 4-6. The sorption percentage of Eu(III) is lowest at pH 6-8, suggesting that a considerable amount of Eu(III) presents as $Eu^{3+}$(HS) complexes in aqueous phase (5). At high pH, sorption of Eu(III) on attapulgite rises again. This is attributed that Eu(III) begins to form carbonate complexes, and the complexes enhance Eu(III) sorption on attapulgite surfaces. However, the sorption of Eu(III) does not reach the level achieved in the system free from FA or HA. This could be due to either residual $Eu^{3+}$(HS) complexes remaining in solution, thus reducing $Eu(CO_3)_3^{3-}$ concentrations available for sorption, or the residual FA or HA on the mineral surfaces enhancing electrostatic repulsion of $Eu(CO_3)_3^{3-}$, or blocking potential sorption sites on the mineral surfaces. Similar phenomena were also found in the sorption of other actinides and lanthanides (5, 20,

*21*). The Laser-introduced fluorescence (LIF) clearly showed that Cm(III) was adsorbed as Cm(III)-fulvate complex in the FA-montmorillonite hybrids at pH below 5, whereas other inorganic species (i.e., carbonate complexes) was the main species at high pH values (*5*).

**Surface complexation model.** The sorption data of Eu(III) on attapulgite in Figure 1A are fitted using constant capacity model (CCM) with the aid of FITEQL 3.2 code. The main sorption species are $\equiv S^wOHEu^{3+}$ and $\equiv X_3Eu^0$ in 0.01 mol/L NaClO$_4$ solution, whereas $\equiv X_3Eu^0$, $\equiv S^wOHEu^{3+}$ and $\equiv S^sOEu(OH)_2^0$ are the dominated species in 0.1 mol/L NaClO$_4$ solution. At high pH values, the main species of adsorbed Eu(III) on attapulgite is $\equiv S^sOEu(OH)_2^0$ in 0.1 mol/L NaClO$_4$ (Figure SI-9 and Table 1). The species of adsorbed Eu(III) can be interpreted as: (1) the strong sites ($\equiv S^sOH$) are formed to $\equiv S^sOH_2^+$ species at low pH. There is very strong electrostatic exclusion between $\equiv S^sOH_2^+$ and $Eu^{3+}$; (2) the $\equiv S^wOH$ site concentration is the highest at low pH; (3) the dominant surface species of attapulgite are $\equiv S^sO^-$ and $\equiv S^wO^-$ at high pH values. However, the complexation ability of strong sites ($\equiv S^sO^-$) with Eu(III) is much stronger than that of weak sites ($\equiv S^wO^-$) (*22*). The main sorption species of Eu(III) on goethite were $\equiv S^wOHEu^{3+}$ and $\equiv S^sOEu(OH)_2$ in no background electrolyte, whereas $\equiv S^wOHEuCl^{2+}$ and $\equiv S^sOEu(OH)_2$ prevailed in sodium chloride, and $\equiv S^wOHEu(NO_3)^{2+}$ and $\equiv S^sOEu(OH)_2$ presented in the nitrate media (*1*). The Eu(III) sorption on Na- and Ca-montmorillonite was modeled well using cation exchange and the monodentate surface species, $\equiv SOEu^{2+}$, $\equiv SOEuOH^+$, and $\equiv SOEu(OH)_3^-$. However, the same data were almost equally well described by considering bidentate surface complexes, $(\equiv X)_2Eu^+$ and $(\equiv S^wO)_2Eu(OH)_2^-$ *(15)*.

**XPS analysis.** The XPS analysis shows that the bonding energy (B.E.) of Eu $3d_{5/2}$ increases from 1134.94 eV on attapulgite to 1135.01 eV on HA-attapulgite hybrids. The main peak of Eu $3d_{5/2}$ changes about 0.07 eV, which indicates that HA has changed the mechanism or species of Eu(III) sorption on attapulgite. The Eu $3d_{5/2}$ core level region spectrum is fitted by deconvolution. In binary Eu/attapulgite system, two peaks at 1133.89 and 1135.60 eV are achieved; however, three peaks at 1134.41eV, 1134.89 and 1136.34eV are used to fit the Eu $3d_{5/2}$ spectrum well in ternary Eu/HA/attapulgite system (Figure SI-11). According to the surface complexation model results, the peaks in the range of 1133.41 eV and 1134.89 eV may be assigned to the species of $\equiv S^wOHEu^{3+}$ and $\equiv X_3Eu^0$, respectively. The peak at 1136.34 eV corresponds to O=C-O-Eu-O-, which is due to the functional groups of HA *(23)*. The peak at 284.62 eV, which is assigned to C 1s, is enhanced obviously in ternary Eu/HA/attapulgite system (Figure SI-12), suggesting that many functional groups are introduced to the surfaces of attapulgite *(24, 25)*. The curve-fitted C 1s and Eu $3d_{5/2}$ region XPS spectra for the samples are discussed in SI (S3.5, Figures SI-10, SI-11 and SI-12). The XPS analysis indicates that the functional groups of HA contribute to Eu(III) sorption to attapulgite, which enhances Eu(III) sorption at low pH values.

**EXAFS Analysis.** For all samples, an intense adsorption line at 6976.9 eV dominates the X-ray adsorption edge *(26)*. The position of this line shows that Eu is trivalent in all of experimental samples *(7, 27, 28)*.

The EXAFS spectra of reference samples (i.e., $Eu_2O_3$, $Eu(OH)_3(s)$ and Eu(aq)) are shown in Figure SI-13. The results are in agreement with the reported references (29, 30). Detailed results and interpretations of the reference samples are listed in SI (S3.6).

The $k^2$-weighted EXAFS oscillation for Eu-HA complexes in different pH values looks similar to each other (Figure 2). However, these spectra somewhat differ from the spectra of the reference samples. The Fourier transforms (FTs) magnitudes for Eu $L_{III}$-edge EXAFS spectra of Eu in HA solution at different pH values are shown in Figure 2A. The first FTs peak around 1.79 Å arises from the single scattering (SS) of the photoelectron on oxygen atoms in the first coordination sphere. It displays a major peak at R = 1.79 Å, which can be related to the contribution of the nearest O shells at 2.4 Å. Several FTs peaks are significantly difference from pH 1.76 to 9.50: (1) the first peak of sample at pH 1.76 is sharper (smaller full width half maximum) than the first peaks of others at other pH values; (2) the oscillation are obvious different in the range of 1.79 to 3.8 Å, indicating that the environment of central atom Eu has been changed, which is possible to distinguish between the Eu-O distances from the hydration sphere and the Eu-O originating from interactions of Eu and oxygen atoms. Figure 2B shows the experimental (open circles) and model (solid line) Fourier-filtered $k^2\chi(k)$ contribution for the next-nearest backscattering shells at (R+ΔR) distances spanning the [1.3, 2.9 Å] interval. The relative parameters of the coordination number (N), interatomic distance (R), residual factors ($R_f$), and EXAFS Debye-Waller factor ($\sigma^2$) are listed in Table 2. From Table 2, the distances of d (Eu-O) decrease from 2.415 to 2.360 Å with increasing pH from 1.76 to 9.50, and the N values also decrease from 9.94 to 8.56. It can be attributed to the contribution of the hydration sphere Eu-OH$_2$ and the carboxylate groups Eu-O(HA), which is a dodecahedron or mono-crown anti-tetragonal-prism structure at pH = 9.50 (Figure 4a and 4b). The measured distances of 2.415 Å (N = 9.94, $\sigma^2$ = 0.0010 Å$^2$ and R$_f$ = 0.032) at pH 1.76 and 2.395 Å (N = 9.97, $\sigma^2$ = 0.0055 Å$^2$ and R$_f$ = 0.048) at pH 2.85 are attributed to the hydration inner-sphere Eu-OH$_2$, which suggests that HA does not form complexes with Eu$^{3+}$, and Eu$^{3+}$ is a typical mono-crown anti-tetragonal-prism (N = 9) or bi-crown-dodecahedron (N = 10) structure (Figure 4b and 4c). In addition, it is very interesting to notice from Figure 2B that the trough at 2.1 and 5.7 Å$^{-1}$ is regular changed from pH 1.76 to 9.50. The XAFS study of Eu(PAA) complexes indicated that two carboxylate groups with 2-3 (Eu(PAA)) and 4-5 (Eu(PAA)$_{ads}$) water molecules were coordinated to Eu in the first coordination sphere (27). The results of Fe(III) interaction with FA showed that Fe(III) was octahedrally configured with inner-sphere Fe-O interactions at 1.98 - 2.10 Å (6).

Figure 3 shows the FTs and the isolated $k^2$-weighted oscillation of the different addition sequences of Eu/HA/attapulgite (detailed description of the addition sequences is listed in SI). The corresponding parameters are listed in Table 2. The FTs and imaginary parts indicate that the addition sequences have slightly effect on Eu(III) sorption and species on attapulgite. From the Fourier-filtered $k^2\chi(k)$ contribution for the next-nearest backscattering shells, the isolated $k^2$-weighted oscillation at 2.48 Å$^{-1}$ and 4.89 Å$^{-1}$ moves significantly forward to lower $k$ values. From Table 2, the mean bond distances of Eu-O are 2.411 Å (N = 11.91, $\sigma^2$ = 0.0163 Å$^2$) for batch 1, 2.399 Å (N = 11.66, $\sigma^2$ = 0.0163 Å$^2$) for batch 2; 2.321 Å (N = 8.866, $\sigma^2$ = 0.0053 Å$^2$) for batch 3, and 2.314 Å (N = 8.24, $\sigma^2$ = 0.0016 Å$^2$) for binary system in the absence of HA at pH 4.50. One can see that the mean bond distances of d (Eu-O) and N values decrease slightly from batch 1 to batch 3. The values of d (Eu-O) and N of batches 1 and 2 are very close to each other, whereas those of d (Eu-O) and N of batch 3 and binary Eu/attapulgite system are similar because Eu(III) is adsorbed on attapulgite firstly in batch 3, which suggests that the configuration of Eu(III) is different (Figure 4a, 4b and 4d). The results indicate that the mechanism and species of Eu(III) sorption to attapulgite has been changed in the different addition sequences. The EXAFS analysis indicates that the influence of addition sequences of HA/Eu(III) on Eu(III) sorption to attapulgite is quite different, and the species are dependent on the different complexation sequences. The results indicate that the presence of natural organic matter and pH values have significant influence on the species of Eu(III), which is crucial to the chemical behavior of Eu(III) in the environment.

**Acknowledgements**

Financial support from National Natural Science Foundation of China (20677058, 20871062, j0630962) and the 973 project (2007CB936602) from MOST of China are acknowledged. The authors greatly acknowledge Dr. Bo He and Dr. Zhi Xie (NSRL, USTC) for helpful technical assistance of EXAFS experiments. We also express our thanks to Prof. B. Grambow (SUBATECH, France) for favorable discussions to improve the quality of the paper.

**Supporting Information Available**

Characterization of attapulgite; Relative species of Eu(III); Surface complex speciation repartition diagram of Eu(III) sorption on attapulgite; XPS spectra; EXAFS spectra and first-shell fit of the EXAFS function of reference and sorption samples. This information is available free of charge via the Internet at http://pubs.acs.org.

**Table 1 Sorption Model of Eu(III) to Attapulgite by Using FITEQL 3.2 Code**

| Ionic strength | Formation | LogK | WSOS/DF |
|---|---|---|---|
| 0.1 mol/L NaClO$_4$ | $3 \equiv XNa + Eu^{3+} \leftrightarrow \equiv X_3Eu + 3Na^+$ | 8.13 | 0.013 |
|  | $\equiv S^sOH + Eu^{3+} + 2H_2O \leftrightarrow \equiv S^sOEu(OH)_2^0 + 3H^+$ | -18.3 |  |
|  | $\equiv S^wOH + Eu^{3+} \leftrightarrow \equiv S^wOHEu^{3+}$ | -2.11 |  |
| 0.01 mol/L NaClO$_4$ | $3 \equiv XNa + Eu^{3+} \leftrightarrow \equiv X_3Eu + 3Na^+$ | 8.46 | 0.015 |
|  | $\equiv S^wOH + Eu^{3+} \leftrightarrow \equiv S^wOHEu^{3+}$ | -1.95 |  |

**Table 2 Mean Eu-O Bond Distance in Aqueous Solution of Eu(III) and HA at Different Conditions, $T = 20 \pm 1\ ^{o}C$, $I = 0.01$ mol/L $NaClO_4$**

| sample | d (Eu-O) (Å) | N | $\sigma^2$ (Å$^2$) | $R_f$ | $\Delta E_0$ |
|---|---|---|---|---|---|
| (1) Reference samples | | | | | |
| $Eu_2O_3$ | 2.347(34) | 6.61(23) | 0.0108(4) | 0.030 | 2.81(26) |
| $Eu(OH)_3$ | 2.408(8) | 8.05(23) | 0.0081(1) | 0.022 | 2.15(67) |
| Eu(aq) | 2.418(3) | 9.24(45) | 0.0084(1) | 0.030 | 2.70(15) |
| (2) Mean Eu-O bond distance in aqueous solution of Eu(III) and HA at different pH | | | | | |
| pH 1.76 | 2.415(12) | 9.94(52) | 0.0010(2) | 0.032 | 2.85(25) |
| pH 2.85 | 2.395(40) | 9.97(75) | 0.0055(8) | 0.048 | 2.41(65) |
| pH 4.80 | 2.392(18) | 8.61(80) | 0.0049(4) | 0.052 | 1.67(1.4) |
| pH 7.72 | 2.385(4) | 8.23(32) | 0.0067(1) | 0.025 | 2.13(91) |
| pH 9.50 | 2.360(6) | 8.56(51) | 0.0036(5) | 0.036 | -0.22(55) |
| (3) Mean Eu-O bond distance of Eu(III) at different addition sequence of HA and HA, pH = 4.50, $I = 0.01$ mol/L $NaClO_4$ | | | | | |
| batch 1 | 2.411(27) | 11.91(89) | 0.0163(1) | 0.045 | 2.4(1.3) |
| batch 2 | 2.399(32) | 11.66(1.2) | 0.0163(1) | 0.066 | 1.4(1.51) |
| batch 3 | 2.321(9) | 8.86(62) | 0.0053(32) | 0.062 | 0.4(6) |
| binary system | 2.314(42) | 8.24(54) | 0.0016(4) | 0.043 | -3.5(5) |

d: Interatomic distance, N: Number of neighbor oxygens, $\sigma^2$: Debye-Waller factor, $\Delta E_0$: Energy shift, $R_f$: The residual factor $R_f = \sum_k (k^3 x_{exp} - k^3 x_{calc}) / \sum_k (k^3 x_{exp})$ measures the quality of the model Fourier-filtered contribution ($x_{calc}$) with respect to the experimental contribution ($x_{exp}$).

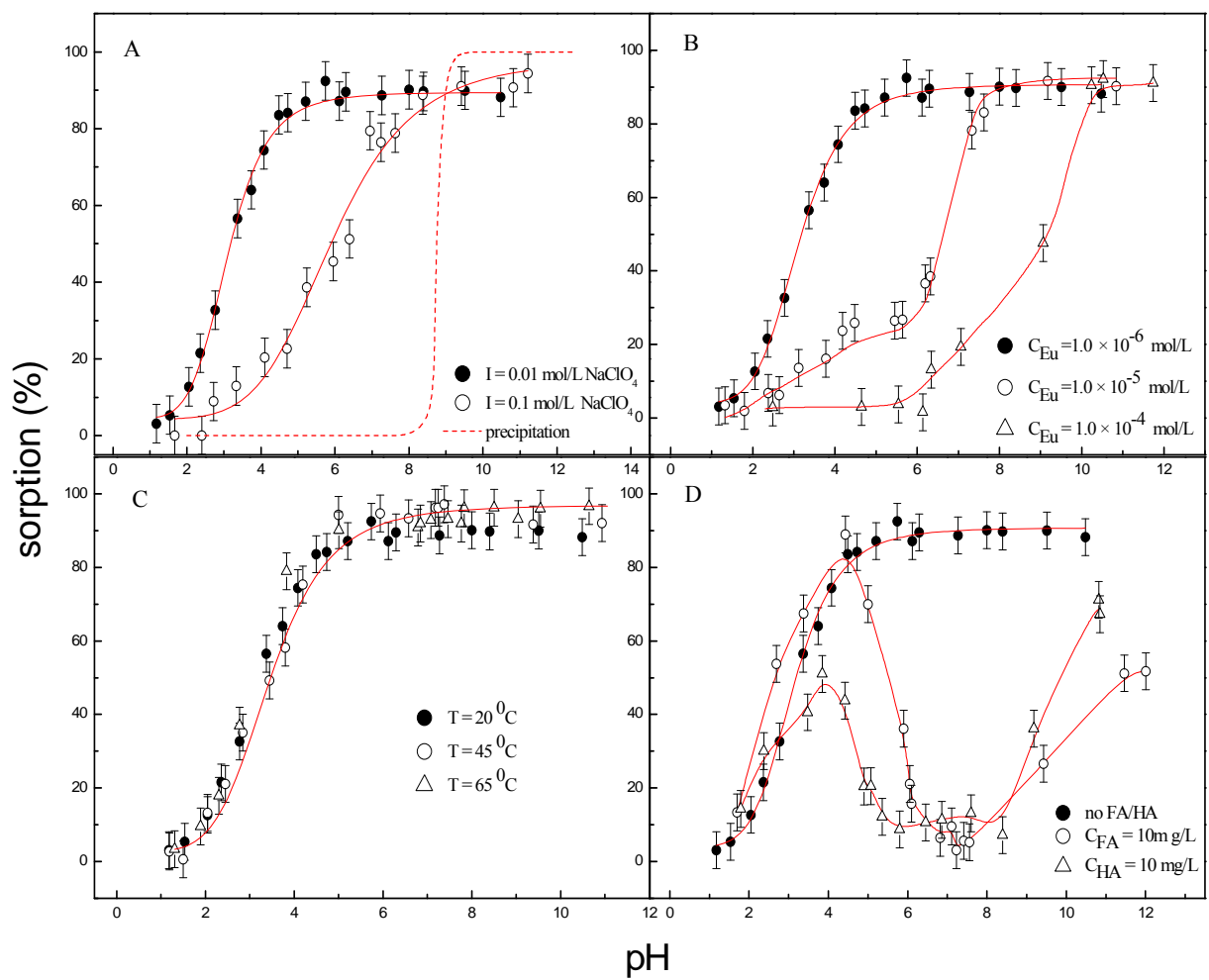

FIGURE 1. Effect of pH, Eu(III) concentration, temperature, ionic strength and FA/HA on Eu(III) sorption to attapulgite. A: Effects of pH and ionic strength on Eu(III) sorption to attapulgite. m/V = 0.1 g/L; $T = 20 \pm 1$ °C; C[Eu(III)]$_{initial}$ = $1.0 \times 10^{-6}$ mol/L; B: Effect of Eu(III) concentration on sorption edge of Eu(III) to attapulgite. m/V = 0.1 g/L; I = 0.01 mol/L NaClO$_4$; $T = 20 \pm 1$ °C; C: The sorption percentage of Eu(III) on attapulgite at different temperatures. m/V = 0.1 g/L; I = 0.01 mol/L NaClO$_4$; C[Eu(III)]$_{initial}$ = $1.0 \times 10^{-6}$ mol/L; D: Effect of FA/HA on Eu sorption to attapulgite at different pH values. m/V = 0.1 g/L; $T = 20 \pm 1$ °C; I = 0.01 mol/L NaClO$_4$; C[FA/HA] = 10 mg/L; C[Eu(III)]$_{initial}$ = $1.0 \times 10^{-6}$ mol/L.

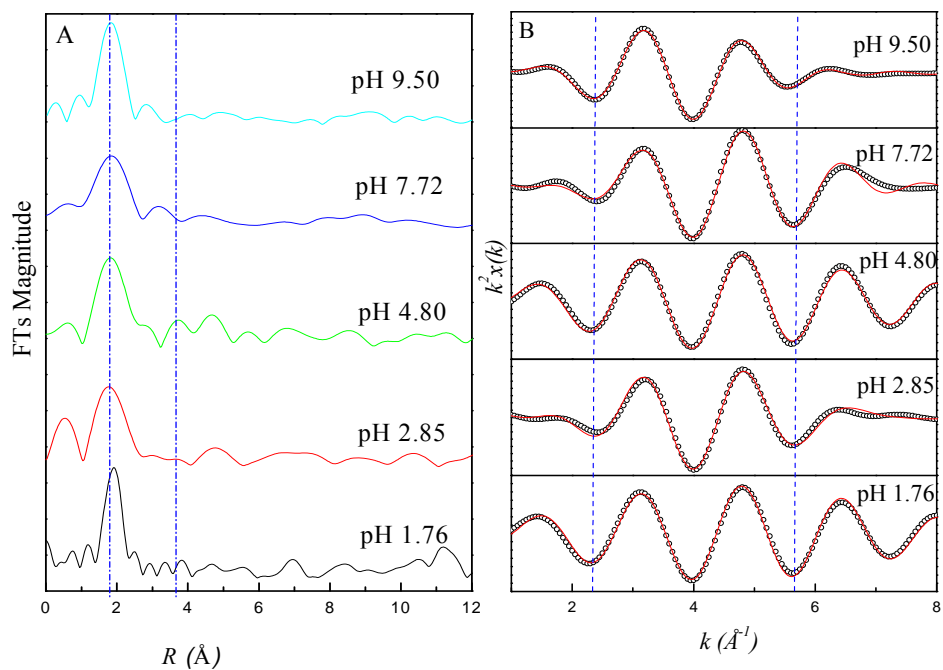

FIGURE 2. The corresponding Fourier Transforms (A) and the first-shell fit of the EXAFS function of $k^2$-weighted (B) of the binary Eu/HA system at different pH values. A: Eu $L_{III}$-edge EXAFS spectra of FTs magnitudes at pH 1.76, pH 2.85, pH 4.80, pH 7.72 and pH 9.50 solutions; B: Experimental (open circles) and model (solid line) Fourier-filtered $k^2\chi(k)$ contribution for the next-nearest backscattering shells at R distances spanning the [1.3, 2.9 Å] interval.

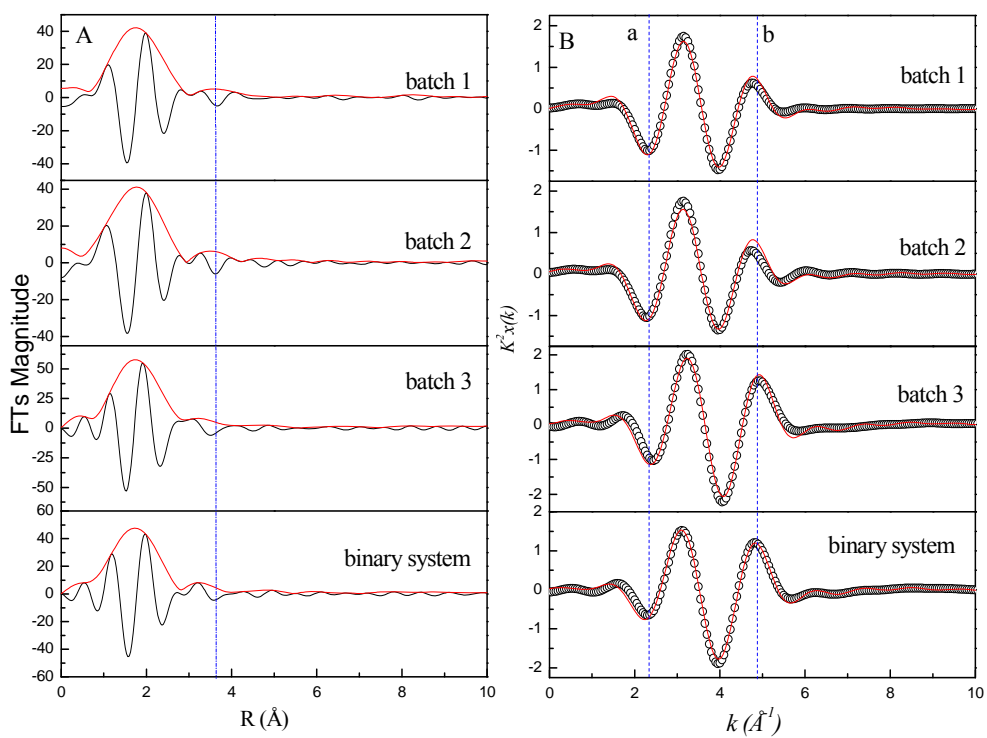

FIGURE 3. The corresponding Fourier transforms (A) and $k^2$-weighted EXAFS (B) of the different addition sequences of ternary Eu/HA/attapulgite systems. A: Eu $L_{III}$-edge EXAFS spectra of FTs magnitudes and imaginary parts, B: Experimental (open circles) and model (solid line) Fourier-filtered $k^2\chi(k)$ contribution for the next-nearest backscattering shells at R distances spanning the [0.6, 2.9 Å] interval.

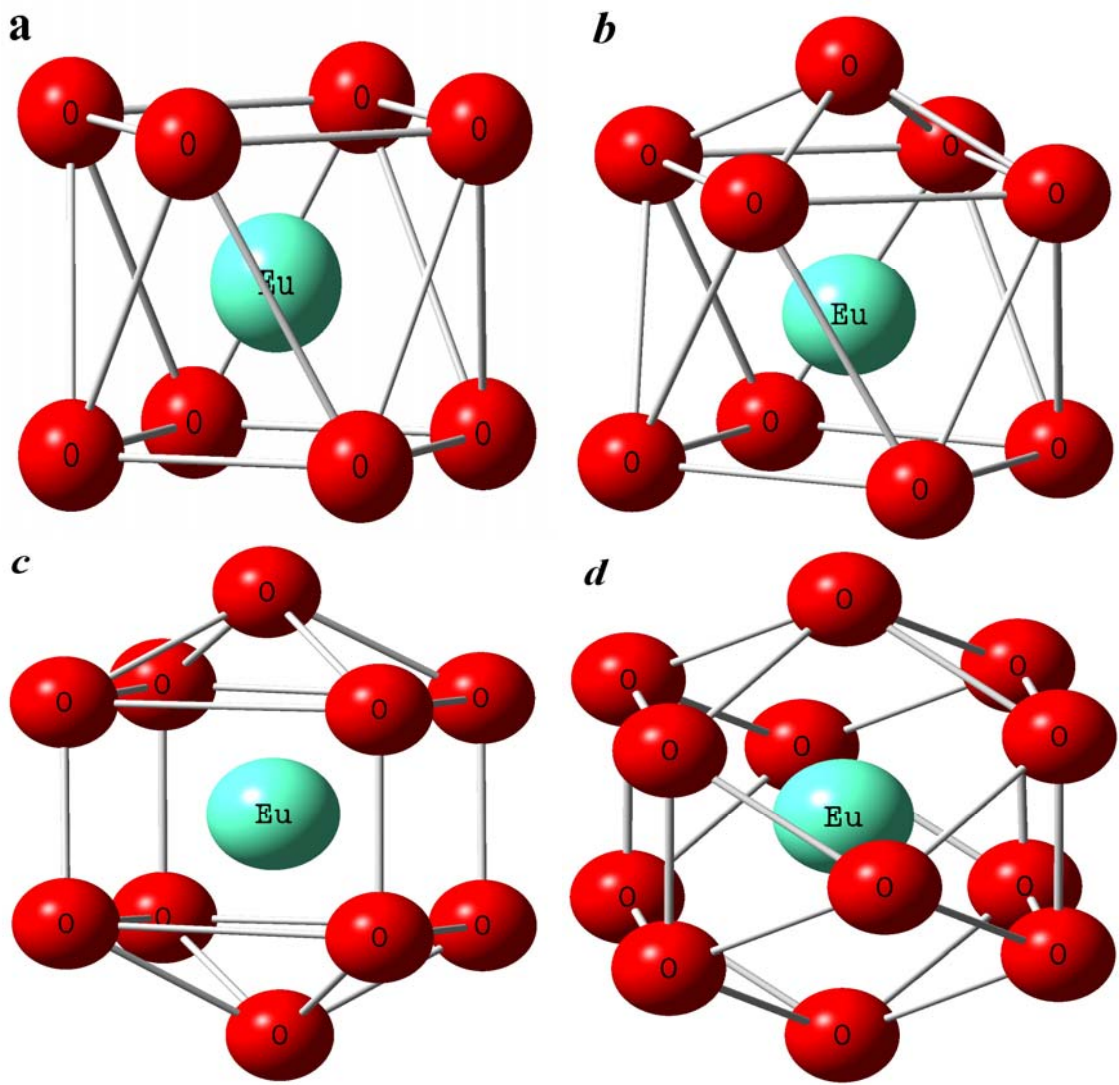

FIGURE 4. The different imaginary Eu(III) structures according to EXAFS results of the first order coordination shells. (a): dodecahedron; (b): mono-crown anti-tetragonal-prism; (c): bi-crown-dodecahedron; (d): icosahedron.

**BRIEF**: The structure of Eu(III) adsorbed on attapulgite is influenced by the different sorption sequences of Eu(III) and HA to attapulgite.